\documentclass[10pt, conference, letterpaper]{ IEEEtran}
\usepackage{cite}
\usepackage{amsmath,amssymb,amsfonts}
\usepackage{algorithmic}
\usepackage{graphicx}
\usepackage{textcomp}
\usepackage{xcolor}
\usepackage[ruled,vlined]{algorithm2e}
\usepackage{array}
\usepackage[OT1]{fontenc} 
\usepackage{stfloats}
\usepackage{url}
\usepackage{verbatim}
\usepackage{wasysym}
\usepackage{multirow}
\usepackage{subcaption}
\usepackage{enumitem}
\usepackage{amsmath}
\usepackage{booktabs}
\usepackage{makecell}
\usepackage[export]{adjustbox}
\usepackage[most]{tcolorbox}
\def\BibTeX{{\rm B\kern-.05em{\sc i\kern-.025em b}\kern-.08em
    T\kern-.1667em\lower.7ex\hbox{E}\kern-.125emX}}

\definecolor{knowledgefill}{RGB}{188,207,225}
\definecolor{knowledgeframe}{RGB}{70,70,70}
\newcommand{\knowledgebox}[2]{%
  \begin{tcolorbox}[
    colback=knowledgefill,
    colframe=knowledgeframe,
    arc=3pt,
    boxrule=0.8pt,
    left=5pt,
    right=5pt,
    top=3pt,
    bottom=3pt,
    width=\linewidth,
    center,
    before skip=6pt,
    after skip=6pt]
    \textbf{#1} #2
  \end{tcolorbox}
}

\newcommand{\pa}{\textsf{PA3 }}
\newcommand{\cmark}{\CIRCLE}
\newcommand{\halfcmark}{\LEFTcircle}
\newcommand{\xmark}{\Circle}

\DontPrintSemicolon 

\SetCommentSty{mycommfont}

\graphicspath{{img/}}
\DeclareGraphicsExtensions{.pdf,.jpeg,.png}
  
\begin{document}

\title{Mitigating Proxy-Induced Traffic Drift in Website Fingerprinting via Model-Agnostic Traffic Tailoring}

\author{\IEEEauthorblockN{Linxiao Yu\IEEEauthorrefmark{1}\IEEEauthorrefmark{2},
Tianyu Cui\IEEEauthorrefmark{2},
Xinhao Deng\IEEEauthorrefmark{3},
Yuqi Qing\IEEEauthorrefmark{3},
Jun Tao\IEEEauthorrefmark{1}\IEEEauthorrefmark{2},
Ke Xu\IEEEauthorrefmark{4}\IEEEauthorrefmark{2},
Qi Li\IEEEauthorrefmark{3}\IEEEauthorrefmark{2}}
\IEEEauthorblockA{\IEEEauthorrefmark{1}School of Cyber Science and Engineering, Southeast University, \IEEEauthorrefmark{2}Zhongguancun Laboratory}
\IEEEauthorblockA{\IEEEauthorrefmark{3}Institute for Network Sciences and Cyberspace, Tsinghua University}
\IEEEauthorblockA{\IEEEauthorrefmark{4}Department of Computer Science and Technology, Tsinghua University}
\IEEEauthorblockA{Emails: \{yulinxiaoybbb, juntao\}@seu.edu.cn, cuity@zgclab.edu.cn, qyq21@mails.tsinghua.edu.cn}
\IEEEauthorblockA{\{dengxinhao, xuke, qli01\}@tsinghua.edu.cn}
}

\markboth{Journal of \LaTeX\ Class Files,~Vol.~14, No.~8, August~2021}%
{Yu \MakeLowercase{\textit{et al.}}: Mitigating Proxy-Induced Traffic Drift in Website Fingerprinting}

\IEEEpubid{0000--0000/00\$00.00~\copyright~2021 IEEE}

\maketitle

\begin{abstract}
Website fingerprinting (WF) based on deep learning can effectively identify websites from encrypted traffic. However, users often rely on proxy protocols to bypass censorship, and the diversity of these protocols poses a major challenge, as WF models trained on traffic from one set of protocols perform poorly when evaluated on that from unseen protocols. We attribute this issue to proxy-induced feature drift, where traffic patterns of the same website vary with the proxy protocol, leading to discrepancies that WF models fail to capture and severe performance degradation. To tackle this issue, we propose \textsf{PA3}, a model-agnostic preprocessing framework to analyze and mitigate the proxy-induced drift. \pa first fingerprints the protocol-specific drift. These fingerprints are then used to tailor the proxied traffic for feature alignment, which mitigates the drift and considerably improves the generalization of WF models on traffic from unseen protocols. Extensive evaluations demonstrate that \pa substantially enhances generalization on unseen protocols with an average improvement of 0.12 in F1-score (roughly 27\% relative), achieving up to a 0.41 absolute gain across models, which narrows the performance gap introduced by the drift. In the best case, \pa enables WF models to obtain F1-scores above 0.96 on traffic from unseen protocols.
\end{abstract}


\section{Introduction}
\label{sec:intro}

\IEEEPARstart{T}he recent decades have witnessed the regulation and censorship on the Internet across multiple countries, which aim to monitor or block user activities deemed improper~\cite{size_info,rtt_info}, such as sending sensitive content or accessing restricted websites. Common censorship techniques include website blocking~\cite{GFWeb, quic_censor} and DNS redirection~\cite{dns_spoofing}. In reaction to the surveillance, proxy techniques have been developed to bypass such restrictions~\cite{rtt_info}. Users typically resort to a proxy client and server to encapsulate original traffic into proxied traffic. A \textit{proxy protocol} specifies how the client communicates with the server, often obfuscating the traffic to evade inspection. Despite the obfuscation, the proxied traffic still leaks informative features.  For example, censors can locate the TLS handshake prologue via packet sizes~\cite{size_info}, or exploit RTT discrepancies across layers~\cite{rtt_info} to detect proxy usage. Additionally, active probing techniques allow censors to even identify the specific proxy protocol in use~\cite{active_probe,firewall,probe_resist}.

To analyze user behavior at a finer granularity, the adversaries have proposed \textit{Website Fingerprinting} (WF)~\cite{DF18}, which infers the visited websites from encrypted traffic by leveraging traffic pattern discrepancies. With the rise of deep learning (DL), numerous studies~\cite{NetCLR23, ARES23, DOM25, EarlyStage24, Scale24, Contrastive24} have demonstrated over 95\% accuracy in website identification. Recent works have also shown that WF can remain effective on the proxied traffic~\cite{ContextAware21, ContextAware24, CrossEnvironmental25}. 

Most existing WF works implicitly rely on a \textit{protocol consistency} assumption that the same set of proxy protocols appears in both training and testing traffic datasets. In practice, however, users might freely switch among diverse proxy protocols (e.g., VMess, Shadowsocks, Trojan), and a model trained on one protocol may encounter traffic from a completely different one at test time. We find that this mismatch causes severe performance degradation. Models tend to learn and compress protocol-specific communication patterns (e.g., handshake sequence pattern) that are shared across all training samples, since such commonality provides little discriminative information for distinguishing websites~\cite{shwartz2017opening, OnInfoBottleneck}. When the testing traffic uses an unseen protocol with different such patterns, the learned compression no longer applies, leading the model to incorrectly process the test features and suffer substantial accuracy loss.


Our key observation is that proxy-induced drift originates from the extra bytes inserted by proxy protocols for encryption and proxy-layer communication. These bytes are irrelevant to website identity, yet they perturb the packet sizes and segmentation patterns observed by WF models. Moreover, proxy software handles each connection independently, which means that the drift is introduced at the flow level, while a website trace is only a mixture of multiple concurrent flows. Therefore, the intuition is to remove these extraneous bytes from each flow to align the traffic in the testing set with that in the training set. In other words, we tailor the traffic to make the protocol consistency assumption feasible.
\IEEEpubidadjcol

Directly removing these irrelevant bytes is infeasible in practice because the user traffic available during the actual WF attack is encrypted. We therefore propose \textsf{PA3}\footnote{The name is inspired by the well-known JA3 fingerprinting method~\cite{althouse2017ja3}.} to remove the packets carrying such bytes to implicitly achieve byte-level alignment. \pa works with a two-phase strategy: \textit{(i) Structural fingerprinting of proxied traffic}: It generates the decrypted probe traffic for each proxy protocol in a controlled environment. It then performs a payload-aware comparison of same-website traffic across protocols and records the discrepant packets as protocol-specific drift fingerprints. \textit{(ii) Traffic tailoring for drift mitigation}: \pa compiles model-agnostic rules based on payload-agnostic packet metadata from fingerprints to tailor encrypted traffic, thereby indirectly removing irrelevant bytes. When performing a WF attack, the adversary eavesdrops the user traffic, where decryption is no longer available. She applies the tailoring rules for traffic feature alignment, which effectively mitigates the proxy-layer distortions to improve the WF generalization on traffic from unseen protocols. Our main contributions are threefold.
\begin{itemize}[leftmargin=*,nosep]
    \item We conduct the \textit{first} study on the ground truth of the proxy-induced drift in a payload-aware way, which reveals that the proxies affect traffic characteristics heterogeneously at each stage of the proxied website visit lifecycle;
    \item We propose model-agnostic methods to tailor the fully encrypted proxied traffic for proxy-induced drift mitigation. Crucially, they operate entirely at the preprocessing stage, requiring no changes to the model architecture and thus enabling plug-and-play integration with existing WF models; 
    \item We collect a proxied website traffic dataset using commercial proxy services, and extensive evaluations demonstrate that \pa substantially outperforms existing drift-mitigating methods. Across 6 state-of-the-art WF models, \pa improves the F1-score by 0.12 on average (27\% relative) over the unmitigated baseline and achieves up to a 0.41 absolute gain. In the best case, the WF model obtains F1-scores above $0.96$ on traces from unseen proxy protocols.
\end{itemize}

The source code of \pa and all the datasets are available at https://github.com/wesly2000/PA3.

\section{Related Work}
\label{sec:related_work}
\begin{table*}[ht]
    \centering
    {\fontsize{8pt}{10pt}\selectfont}
    \caption{Overview of proxy protocol usage by proxy service providers. \text{\cmark} indicates protocols currently in use, \text{\halfcmark} indicates protocols that were formerly used, and \text{\xmark} indicates protocols that have never been used by the providers.}
    \begin{tabular*}{\textwidth}{@{\extracolsep{\fill}}ccccccccccc}
        \toprule
            & WgetCloud & SS ONE & Youxin Cloud & CLM & STC-Server & RioLU.443 & DoggyGo & Nexitally & Erwan Cloud & JustMySocks \\
            \midrule
Trojan       & \cmark     & \xmark & \cmark & \halfcmark & \xmark & \cmark     & \xmark & \halfcmark & \xmark & \xmark \\
VMess        & \halfcmark & \cmark & \xmark & \xmark     & \cmark & \xmark     & \xmark & \xmark     & \cmark & \cmark \\
Shadowsocks  & \halfcmark & \cmark & \xmark & \cmark     & \xmark & \cmark     & \cmark & \cmark     & \cmark & \cmark \\
VLess        & \xmark     & \xmark & \xmark & \xmark     & \xmark & \cmark     & \xmark & \xmark     & \xmark & \xmark \\
hy2          & \xmark     & \xmark & \xmark & \xmark     & \xmark & \halfcmark & \cmark & \xmark     & \xmark & \xmark \\
ShadowsocksR & \halfcmark & \xmark & \xmark & \xmark     & \cmark & \xmark     & \xmark & \xmark     & \xmark & \xmark \\
        \bottomrule
    \end{tabular*}
    \label{table:proxy_protocols}
\end{table*}

\noindent\textbf{Website Fingerprinting Attack.} Website fingerprinting captures the traffic feature discrepancies, which encode the uniqueness of website resources, e.g., images and videos, to classify the websites even when the traffic is encrypted~\cite{ETBERT22}. With the flourishing of deep learning (DL), researchers began to adopt DL models in WF analysis~\cite{DF18, AWF18}, which achieved significant success in classification accuracy and scale~\cite{Scale24}. Previous works have demonstrated the strength of DL-based WF, which could precisely classify the websites from HTTPS~\cite{GNN24,ETBERT22,HOLMES25,cebere2024understanding} or Tor traffic~\cite{ARES23,NetCLR23,Scale24,EarlyStage24}, by automatically extracting high-level representations from simple traffic features like packet direction. Even when obfuscation is applied, recent WF works show considerable potential in breaking these safeguards~\cite{RF23,DontSpeak24}.

\noindent\textbf{Proxy-Based Censorship Evasion.} Internet censorship techniques, such as website blocking~\cite{GFWeb, quic_censor} and DNS redirection~\cite{dns_spoofing}, have motivated the development of proxy techniques that encapsulate user traffic to evade inspection~\cite{rtt_info}. A variety of proxy protocols have emerged, including Trojan, VMess (V2Ray), Shadowsocks, VLess, and Hysteria~2 (hy2). We survey 10 popular commercial proxy service providers in Table~\ref{table:proxy_protocols}, and select the three most commonly deployed protocols---VMess, Shadowsocks, and Trojan---for this work. Despite the obfuscation these protocols provide, censors can still detect proxy usage~\cite{size_info,rtt_info,active_probe,firewall,probe_resist}. Assuming the role of the censors, recent works start to apply WF techniques on proxied traffic for a more fine-grained traffic analysis, i.e., to identify the specific websites requested with proxy protocols~\cite{ContextAware21, ContextAware24, CrossEnvironmental25}.

\noindent\textbf{Traffic Drift in WF.} WF performance can deteriorate because of changes in network conditions~\cite{rosetta,proteus2026}, HTTP versions~\cite{HOLMES25}, website content~\cite{ARES23}, etc. For proxied traffic, the drift comes from the variety of the proxy protocols used, where the protocol consistency assumption is often violated. To tackle this issue, \cite{CrossEnvironmental25} proposes the first framework to select the optimal feature set that is robust to this drift. However, its features depend on both the selected websites and access to protocols in the testing dataset, producing substantial variation across common-set selections. Moreover, it means that the features used in WF are not website-independent, limiting its generalizability. In contrast, \pa learns website-independent protocol characteristics and does not require WF models to access traces from protocols in the testing dataset.


\section{Threat Model}
\label{sec:problem}


\begin{figure}[t]
    \centering
    \includegraphics[width=0.4\textwidth]{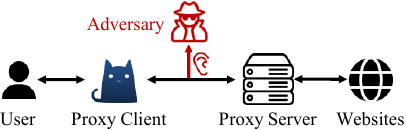}
    \caption{The threat model for fingerprinting proxied traffic.}
    \label{fig:threat}
\end{figure}
Given a set of monitored websites $\mathcal{W}$, users may access them using protocols from the set $P = \{P_1, ..., P_k,\allowbreak P_{k+1}, ..., P_{k+r}\}$. For simplicity, we use \textit{NoProxy} as a protocol to represent the case when no proxy is used. Let $P_{\text{train}} = \{P_1, ..., P_k\}$ be the set of protocols whose traffic is available during training, and $P_{\text{test}} = \{P_{k+1}, ..., P_{k+r}\}$ be the set of protocols only occurring in the testing traffic. Visiting websites in $\mathcal{W}$ using protocols in $P$ generates a traffic dataset $D = \{D_1, ..., D_{k+r}\}$, where $D_i$ represents traffic to $\mathcal{W}$ proxied by protocol $P_i$. Accordingly, we define $D_{\text{train}} = \{D_1, ..., D_k\}$ and $D_{\text{test}} = \{D_{k+1}, ..., D_{k+r}\}$ to represent training and testing datasets, respectively. We assume $P_{\text{train}} \cap P_{\text{test}} = \varnothing$, i.e., the model has no access to $P_{\text{test}}$ traffic during training. 


The adversary trains a WF model on $D_{\text{train}}$ and seeks to improve its generalization to $D_{\text{test}}$. She generates a small amount of controlled, decryptable probe traffic to learn protocol-specific tailoring knowledge. During the attack, however, she can only passively observe encrypted traffic between the proxy client and server. Because the incoming protocol is unknown, she first identifies it using prior protocol knowledge and then applies the corresponding payload-agnostic tailoring rules.

\section{Methodology}
\label{sec:methodology}
\subsection{Overall Framework}
\label{subsec:methodology_overall}

\begin{figure*}[t]
    \centering
    \includegraphics[width=.95\textwidth]{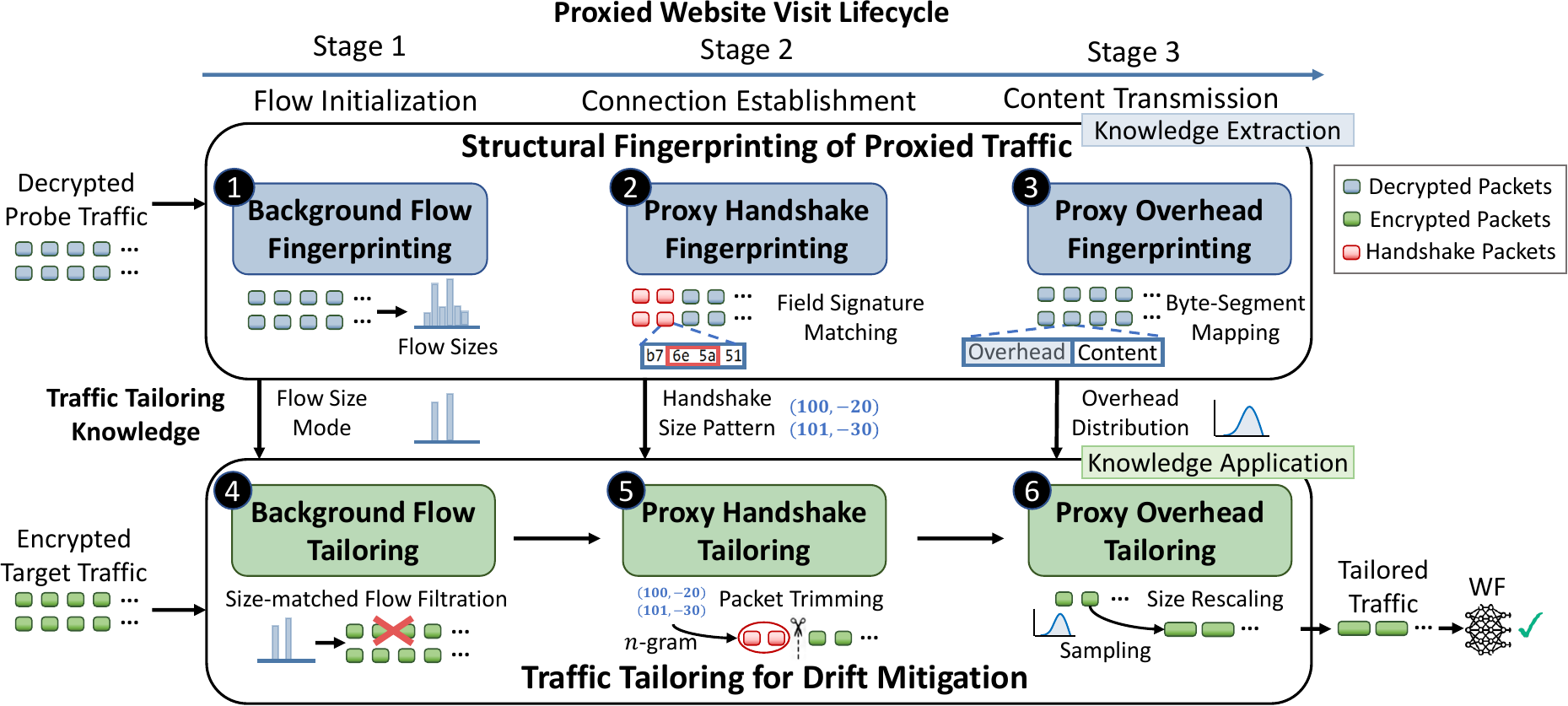}
    \caption{Overview of our methodology. In the knowledge extraction phase, the adversary collects decrypted probe traffic to structurally fingerprint proxied traffic and derive traffic tailoring knowledge. In the knowledge application phase, the adversary applies the learned knowledge to encrypted target traffic and feeds the tailored traces to existing WF models.}
    \label{fig:methodology_overview}
\end{figure*}



We develop \textsf{PA3}, which extracts knowledge from the fingerprints of the protocol-specific drift, and applies the knowledge to tailor the flows for drift mitigation. The intuition behind \pa is that proxy-induced drift originates from the extra bytes introduced for proxy-layer communication. Therefore, we could remove the packets carrying these bytes to indirectly remove them for drift mitigation. We present the framework overview in Fig.~\ref{fig:methodology_overview}. \pa consists of two phases: 

\noindent\textbf{Structural Fingerprinting of Proxied Traffic}: For a given stage in the proxied website visit lifecycle, the packets carrying the extra bytes of different proxy protocols expose protocol-specific patterns. Such regularity allows us to extract stage-specific structural fingerprints for each protocol. The proxy-layer encryption hides the semantic information of each packet. Therefore, we collect a small amount of controlled probe traffic for each proxy protocol, which is decrypted and parsed using the toolchain we developed to build the ground truth on which packets carry the extra bytes. The extracted fingerprints are converted to \textit{traffic tailoring knowledge} which is applicable when the traffic is encrypted.

\noindent\textbf{Traffic Tailoring for Drift Mitigation}: In this phase, \pa applies the traffic tailoring knowledge extracted from the previous phase to encrypted target traffic, where payload inspection is no longer available. \pa scans through each flow to compute the likelihood that the packet sequence matches the patterns in the knowledge. Once the likelihood exceeds some threshold, \pa marks these packets as the ones that carry extra bytes and strips them from the flow. The drift would be mitigated after all flows are tailored, and the WF model would obtain better generalization on the tailored traffic.

\subsection{Structural Fingerprinting of Proxied Traffic}
\label{subsec:structural_fingerprinting}
Structural fingerprinting explains how each lifecycle stage is distorted by the proxy layer. We generate a small amount of controlled probe traffic, record the decryption keys, and compare flows to the same hostname across proxy protocols. This payload-aware comparison isolates proxy-induced effects from ordinary noise such as network congestion, retransmissions, and flow interleaving. The goal is not to use decrypted information during the real WF attack, but to turn the observations into payload-agnostic knowledge that can later be applied to encrypted traces.

\begin{figure}
    \centering
    \includegraphics[width=0.48\textwidth]{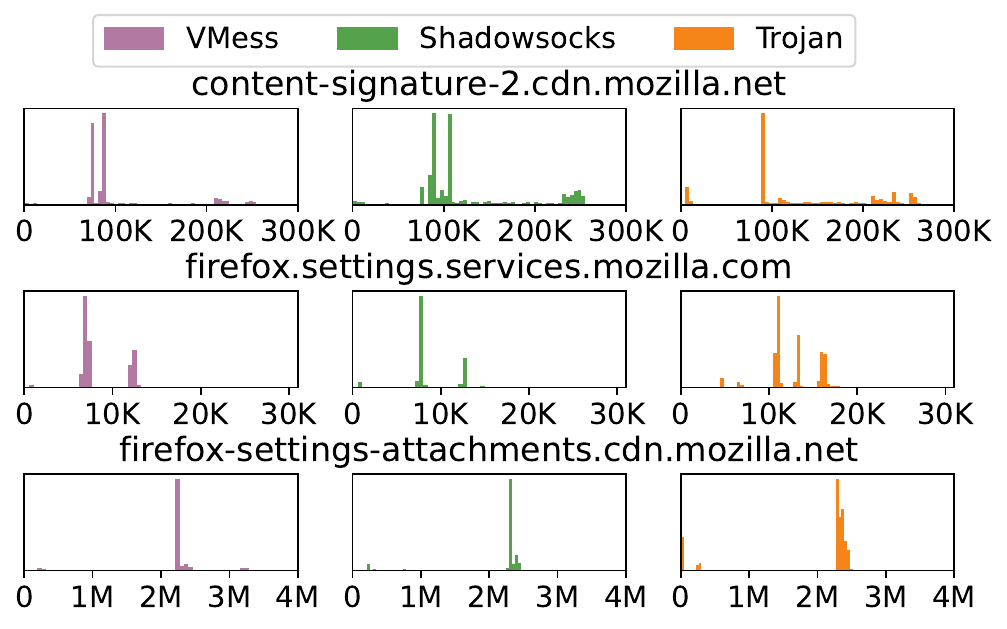}
    \caption{BIH flow size statistics. Each row shows the size histograms of the same hostname under different proxy protocols.}
    \label{fig:browser_noise}
\end{figure}

\begin{figure}
    \includegraphics[width=0.5\textwidth, center]{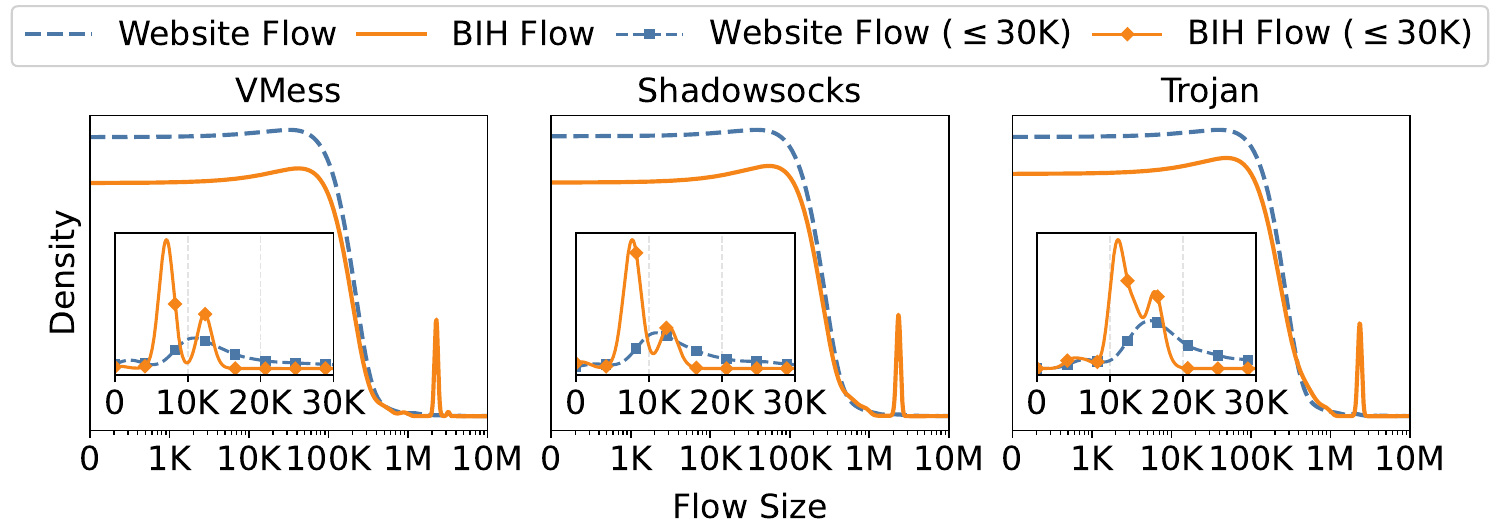}
    \caption{BIH and website flow size distributions. Most of the website flows have sizes between $0\sim30$ KB, and the zoomed views show that the BIH flows have different modes from the website flows within this range.}
    \label{fig:browser_noise_2}
\end{figure}

\noindent\textbf{Background Flow Fingerprinting.} When visiting a website, the browser determines not only the hostnames associated with the target website, but also a set of browser-specific service domains used for auto-update checks, security verification, telemetry, and similar tasks. We refer to these background domains as \textit{Browser Intrinsic Hostnames} (BIHs). For example, Firefox periodically requests \texttt{content-signature-2.\allowbreak cdn.mozilla.net} to retrieve certificate-chain information for Mozilla service data~\cite{content_signature}. Although BIH flows carry no website-discriminative information, proxy-layer overhead shifts their sizes in protocol-specific ways, turning them into protocol-dependent noise that exacerbates cross-protocol feature drift. We therefore need to identify and filter BIH flows.

Because BIH flows perform stable background tasks, their sizes concentrate around protocol-specific modes. Fig.~\ref{fig:browser_noise} shows that these modes form sharp peaks, which provide fingerprints for identifying background flows and enable size-matched flow filtering. We also compare BIH flow sizes with website flow sizes. Fig.~\ref{fig:browser_noise_2} shows that many website flows fall within the $0\sim30$KB range, where BIH flows may also appear. Therefore, background flow tailoring must balance BIH removal against preservation of website flows. BIH flows at larger sizes overlap less with website flows and are less risky to filter.

\knowledgebox{Knowledge I:}{Background flows in the proxied traffic form protocol-specific flow-size modes, enabling their identification and filtering through size matching.}

\noindent\textbf{Proxy Handshake Fingerprinting.} Before connecting to the target web server, the proxy client first establishes communication with the proxy server, and the inner TLS handshake with the web server follows. This delays the TLS handshake by inserting a proxy handshake packet sequence before it. Since each proxy protocol follows its own handshake procedure, the delay is protocol-specific.

\begin{figure}[t]
    \centering
    \includegraphics[width=0.48\textwidth]{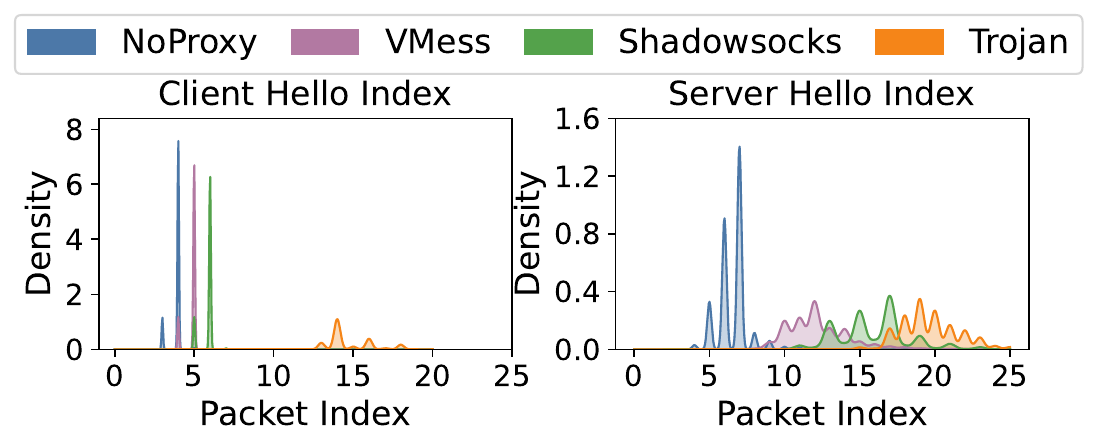}
    \caption{Distribution of intra-flow packet indices of TLS Client Hello and Server Hello.}
    \label{fig:packet_index}
\end{figure}

To quantify this effect, we record the intra-flow packet indices of TLS Client Hello and Server Hello across flows. Fig.~\ref{fig:packet_index} shows how proxy protocols delay both packets compared with the NoProxy case, and that the indices concentrate around protocol-specific modes. Client Hello is especially stable, while Server Hello has slightly more variation because that other packets may appear between Client Hello and Server Hello, like the delayed TCP ACK~\cite{rfc1122} packets triggered by the proxy handshake packets. This concentration indicates that proxy handshake packets form a consecutive anomalous subsequence near the beginning of each flow. The structural regularity of these packet sequences makes them suitable for sequence-based fingerprinting and later packet trimming.

\knowledgebox{Knowledge II:}{Proxy handshakes form protocol-specific directional packet-size patterns near the beginning of each flow, which can guide sequence-based packet trimming.}

\noindent\textbf{Proxy Overhead Fingerprinting.}
After connection establishment, website resources are carried through the proxy tunnel. The proxy layer appends headers and encryption tags to proxied data units, so the payloads passed to TCP are proxy-encapsulated payloads rather than the original TLS payloads. This changes TCP segmentation throughout the flow, which means the drift is distributed rather than localized in removable packets. To fingerprint this distributed overhead, we construct the \textit{Byte-Segment Map} (BSM) structure. $BSM(i)$ returns the intra-flow index of the TCP segment containing the $i$-th HTTP byte in a flow. For the same hostname, comparing the BSMs generated under different protocols reveals how the same upper-layer bytes are segmented into TCP packets. 

\begin{figure}
    \centering
    \includegraphics[width=.5\textwidth, center]{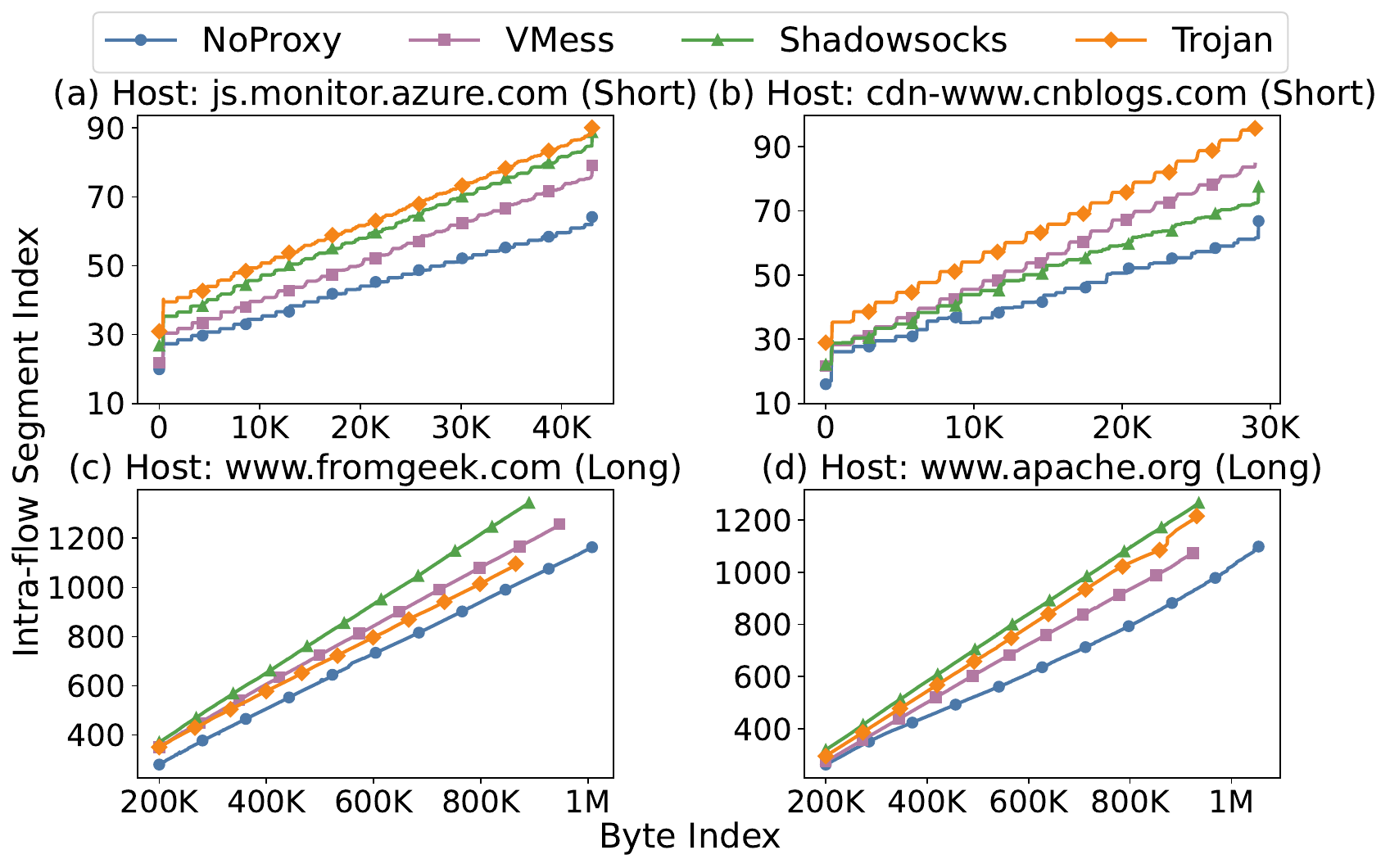}
    \caption{BSMs of HTTP and TCP for several hostnames, which depict how the TCP segmentation of the HTTP contents differs across proxy protocols.}
    \label{fig:byte_segment_map}
\end{figure}

\begin{figure}[t]
    \centering
    \includegraphics[width=0.48\textwidth]{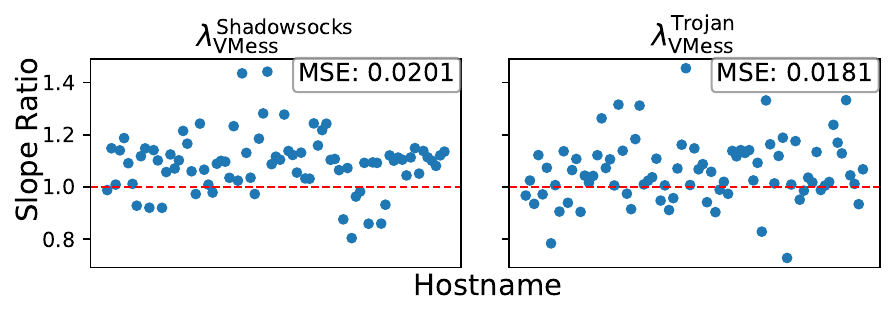}
    \caption{Ratio of slopes between BSMs of different protocols.}
    \label{fig:slope_ratio}
\end{figure}

Fig.~\ref{fig:byte_segment_map} (a)-(b) show the BSMs for short flows to highlight local changes. The starting points of proxied flows are slightly higher than those of NoProxy flows because of the additional proxy handshake. The occasional vertical jumps are caused by TCP ACK packets, which acknowledge the reception of the data but carry no HTTP bytes, thus the byte index remains fixed while the segment index advances. Fig.~\ref{fig:byte_segment_map} (c)-(d) show the BSMs for long flows. This large-scale view reveals that the accumulated proxy overhead during content transmission leads to steeper lines than for NoProxy. These extra bytes are encapsulated into the TCP payloads, so the same HTTP bytes require more TCP segments. The slopes also differ across protocols because of different header formats and encryption strategies. Fig.~\ref{fig:slope_ratio} summarizes the ratio between BSM slopes of protocol pairs. For protocols $P_1$ and $P_2$, we denote the slope ratio as $\lambda^{P_1}_{P_2}$. Although the ratio varies across hostnames because of different chunking strategies~\cite{cloudflare2016dynamic}, the low mean squared error (MSE) suggests that the ratios concentrate in a narrow band. It means that the slope ratio is a useful statistical description of inter-protocol segmentation drift and motivates the burst-size rescaling method during traffic tailoring.

\knowledgebox{Knowledge III:}{Proxy overhead makes each flow inflate at a protocol-specific ratio, which can be used to rescale intra-flow bursts for protocol-specific overhead alignment.}

\subsection{Traffic Tailoring for Drift Mitigation}
\label{subsec:traffic_tailoring}
Proxy protocols distort traffic at each stage of the website visit lifecycle. These distortions all arise from extraneous bytes injected by the proxy layer for encryption and communication. Mitigating these impacts therefore reduces to eliminating these bytes. However, since proxy protocols encrypt their payloads, directly locating and stripping these bytes from the eavesdropped traffic is infeasible. Therefore, we tailor the encrypted target traffic using the knowledge learned from decrypted probe traffic, which leverages only flow sizes, packet positions, and directional packet sizes. This metadata is observable despite encryption, which makes drift mitigation practical in the real-world scenario.

\noindent\textbf{Background Flow Tailoring.}
We empirically remove BIH flows by matching flow sizes against the BIH flow size modes. The main design choice is the filter coverage threshold $\alpha$, i.e., the fraction of the BIH flow-size histogram covered by the filter. A small $\alpha$ leaves many BIH flows unfiltered, whereas a large $\alpha$ risks removing website flows whose sizes overlap with BIH flows. We therefore design the scoring function
\begin{equation}
U(\alpha) = (1+\beta)\frac{C_{filt}\cdot C_{re}}{\beta C_{filt}+C_{re}},
\end{equation}
to select $\alpha$ by balancing the proportions of filtered BIH flows and retained website flows, where $C_{filt}$ is the proportion of BIH flows filtered and $C_{re}$ is the proportion of website flows retained under $\alpha$. The weight $\beta$ controls how the score prioritizes BIH flow removal against website flow preservation. We choose the protocol-specific optimal $\alpha^*$ maximizing $U(\alpha)$.

Since different proxy protocols may yield different $\alpha^*$ values, applying each protocol's optimum independently can leave different amounts of BIH noise in different protocols and introduce a new cross-protocol mismatch. To keep the remaining background flow distribution aligned, we use the shared threshold $\alpha_{opt}=\min\{\alpha_1^*,\alpha_2^*,\ldots,\alpha_n^*\}$ across $n$ protocols. Given an $\alpha_{opt}$, we sort the BIH histogram bins from the highest bar to the lowest and include bins in that order until the selected bins occupy $\alpha_{opt}$ of the whole histogram. The resulting union of bins is the filter, and any flow whose size falls into this scope will be removed.

\noindent\textbf{Proxy Handshake Tailoring.}
To detect and trim the anomalous packet sequence introduced during proxy handshake, we represent each flow as a sequence of directional packet sizes and apply an $n$-gram approach~\cite{Siby2019EncryptedDNS} for proxy handshake detection. A sliding window is marked anomalous if it matches patterns learned from proxy handshake sequences in the probe traffic. Consecutive anomalous windows are merged, and the corresponding packets are removed from the flow. Because the proxy handshake region appears near the beginning of a flow, the search can be limited to the initial packets.

Packet sizes range over many possible values, so we discretize directional packet sizes before constructing the $n$-gram vocabulary. Given vocabulary size $v=|V|$, we divide the range $[-r,r]$ into $v$ equal bins of width $\Delta=2r/v$. Packet $p_i$ with direction $d_i\in\{+1,-1\}$ and size $s_i$ is mapped to
\begin{equation}
Q_{v}(p_i) = d_i \cdot \left\lfloor \frac{s_i}{\Delta} \right\rfloor,
\end{equation}
and window $W_i=(p_i,\ldots,p_{i+k})$ is represented as $Q_v(W_i)=\bigl(Q_v(p_i),\ldots,Q_v(p_{i+k})\bigr)$. The vocabulary size controls the trade-off between precision and recall. Smaller vocabularies conflate distinct patterns, while larger vocabularies increase miss rates. We select the optimal vocabulary size by maximizing the mutual-information objective
\begin{equation}
    \begin{aligned}
    v^*
    &= \operatorname*{arg\,max} J(v), \\
    J(v)
    &= I\!\left(Q_v(W); Y\right) - \gamma \log v ,
    \end{aligned}
    \label{eq:mi_objective}
\end{equation}
where $Y\in\{0,1\}$ indicates whether a window is anomalous, and $\gamma$ controls the penalty. After anomalous packet trimming, we perform SEQ-ACK analysis to remove ACK packets triggered by the proxy handshake. The same anomaly dictionaries also support proxy protocol identification. The flow-level protocol is the one whose dictionary matches the handshake pattern of the target flow. The session-level protocol, i.e., the protocol of the complete traffic from a single website visit, is determined by the majority voting of the flows within it.

\noindent\textbf{Proxy Overhead Tailoring.}
Since the proxy headers and encryption tags are co-carried with website content, the overhead during content transmission is dispersed throughout the entire flow, and removing individual packets would discard useful website information. Instead, we treat the overhead as a burst-scaling effect. Let $B_1$ and $B_2$ be burst sizes under protocols $P_1$ and $P_2$ for the same upper-layer content. The BSM slope ratio $\lambda^{P_1}_{P_2}$ approximates the scaling relation
\begin{equation}
\label{eq:burst_scaling}
B_1 \approx \lambda^{P_1}_{P_2} \cdot B_2 .
\end{equation}

\begin{figure}[t]
    \centering
    \includegraphics[width=.49\textwidth, center]{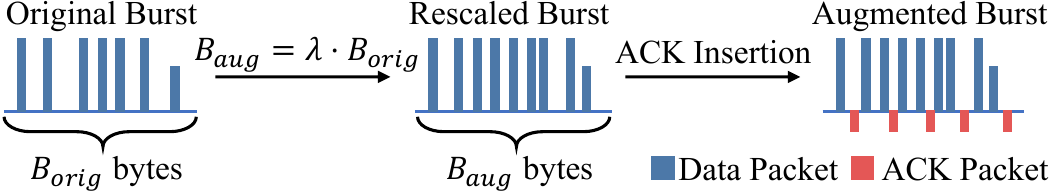}
    \caption{Rescale the burst sizes according to the empirical ratio range, and re-insert ACK according to the empirical ACK frequency CDF.}
    \label{fig:augmentation}
\end{figure}

Fig.~\ref{fig:augmentation} illustrates the augmentation procedure. From the probe traffic, the adversary computes the mean $\mu^{P_1}_{P_2}$ and standard deviation $\sigma^{P_1}_{P_2}$ of the BSM slope ratios for each protocol pair. We model the scaling factor as a Gaussian random variable $\lambda\sim N(\mu^{P_1}_{P_2},\sigma^{P_1}_{P_2})$. During augmentation, the adversary samples $\lambda$ and rescales each original burst $B_{orig}$ to $B_{aug}=\lambda\cdot B_{orig}$. The rescaled burst is reconstructed into a valid packet sequence by filling packets up to TCP Maximum Segment Size (MSS)~\cite{RFCMSS}, with the final packet carrying any remainder. Timestamps are reconstructed by sampling inter-packet delays from the empirical delay distribution of the original flow. Moreover, TCP ACKs often appear after every few data packets. Within a single flow, they divide packets in the same direction into small bursts, and directly rescaling these tiny bursts can create unrealistic packet sequences. We therefore remove ACK packets before burst rescaling, and then re-insert them back after rescaling according to the empirical ACK frequency of the original flow.

\section{Evaluation}
\label{sec:evaluation}
In this section, we evaluate how our mitigation methods improve WF generalization on unseen protocols.

\subsection{Experiment Setup}
\noindent\textbf{Implementation.} We prototype \pa with more than 4,500 lines of C code to enable payload-aware analysis, and with more than 3,500 lines of Python code for drift mitigation. Selenium~\cite{selenium} and the modified Clash Core~\cite{clashCore} are used to automate traffic and key material collection.

\begin{figure}[t]
    \centering
    \includegraphics[width=.48\textwidth, center]{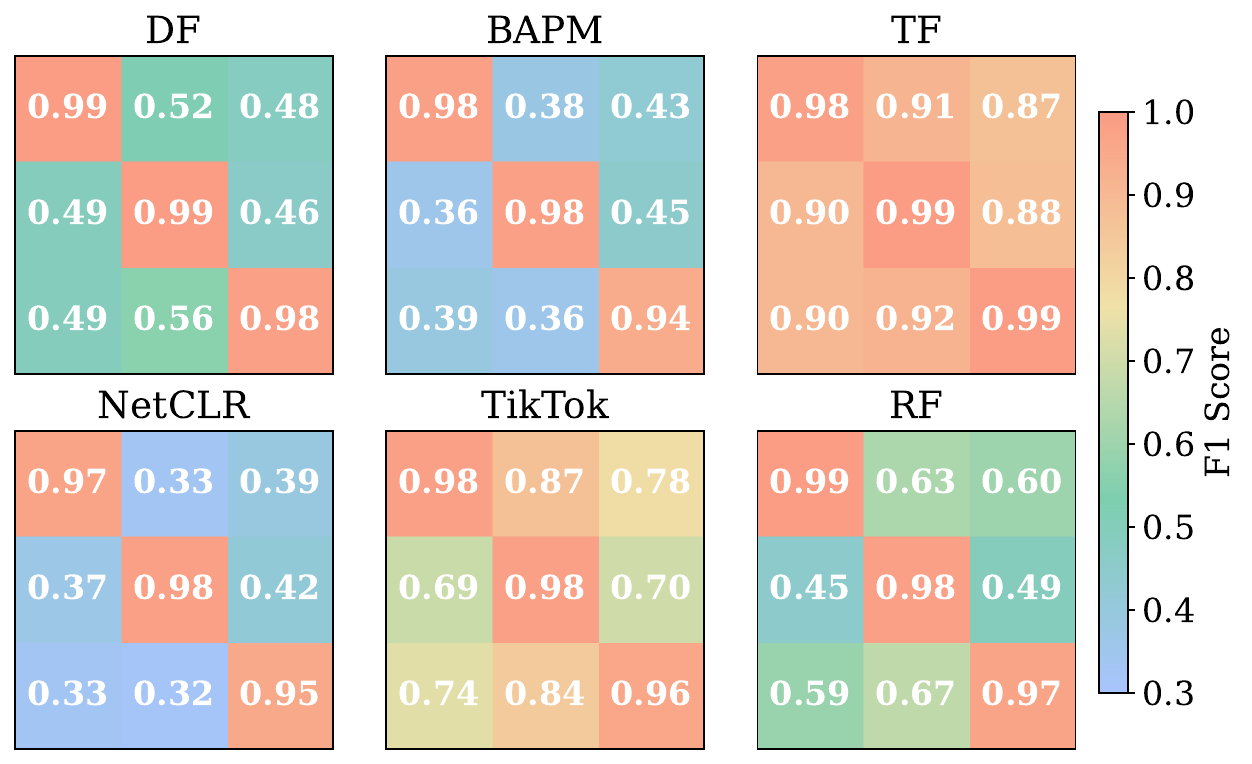}
    \caption{Performance of WF models trained with traffic from protocols in the training dataset ($P_{\text{train}}$), and tested on that from protocols in the testing dataset ($P_{\text{test}}$). When $P_{\text{train}}\cap P_{\text{test}}=\varnothing$ (off-diagonal cases), the performance deteriorates severely.}
    \label{fig:perf_degrade}
\end{figure}

\noindent\textbf{Dataset.} Existing proxied traffic datasets for WF are either heavily processed into statistical features~\cite{ContextAware21,ContextAware24,CrossEnvironmental25}, leaving little space for our payload-aware analysis and flow-level processing, or remain private and difficult to obtain~\cite{rtt_info,size_info}. Recognizing the importance of raw traffic for reproducible traffic analysis research, we collect our proxied traffic dataset with the corresponding key materials to facilitate future studies, which is released along with the \pa repository. Specifically, we collect around 1 TB of traffic data from 80 websites using 4 protocols (NoProxy, VMess, Shadowsocks and Trojan) under diverse network settings. 
\begin{itemize}[leftmargin=*,nosep]
    \item We use commercial proxy services rather than a laboratory deployment to collect more realistic proxied traffic. Moreover, we collect traffic through proxy servers in several regions, including Japan, Singapore, and the United States, to capture geographic diversity.
    \item We deploy 3 client machines across China and the United States. Two clients operate on university campus networks in China, while the third runs as a virtual machine in a U.S. region of a commercial cloud platform.
    \item To understand how the browser version affects the BIH flow size, besides the main dataset we collect another traffic dataset for proxy protocols using Firefox from 10 different versions, i.e., from version 132 (released in October 2024) to version 141 (released in July 2025) individually.
\end{itemize}

\subsection{Mitigating Protocol-specific Distortions}
In this section, we apply our mitigation approaches proposed in Section~\ref{subsec:traffic_tailoring} to improve the model generalization. We consider the F1-score as the main performance metric. In this work, we evaluate 6 state-of-the-art WF models, namely, DF~\cite{DF18}, BAPM~\cite{BAPM21}, TF~\cite{TF19}, NetCLR~\cite{NetCLR23}, TikTok~\cite{TikTok20}, and RF~\cite{RF23}, to validate consistent performance gains across different model architectures.


\noindent\textbf{Performance Degradation on Unseen Protocols.} First, we demonstrate in Fig.~\ref{fig:perf_degrade} how WF performance degrades when the model encounters traffic from unseen protocols. Let $P_{\text{train}}$ and $P_{\text{test}}$ represent the protocols in the training and testing datasets, respectively. Along the diagonal of Fig.~\ref{fig:perf_degrade}, where $P_{\text{train}}=P_{\text{test}}$, all WF models achieve extremely high performance, with F1-scores above 0.94. In contrast, when $P_{\text{train}}\cap P_{\text{test}}=\varnothing$, the models suffer severe performance degradation. For example, the F1-score of DF drops below 0.56. Even relatively drift-tolerant models such as TF and TikTok perform much worse than in the protocol-consistent setting. These results indicate that existing WF models cannot effectively handle the feature drift caused by proxy-protocol inconsistency.

\noindent\textbf{Proxy Protocol Identification.} Before applying protocol-specific tailoring knowledge, \pa first identifies the proxy protocol used by each incoming trace. We leverage the $n$-gram method to identify the proxy protocols. The anomalous patterns that the $n$-gram method learns are extracted from no more than 400 flows per protocol. The identification results are listed in Table~\ref{table:proxy_protocol_identification}. In flow-level identification, fewer than 0.6\% of samples are misclassified for all protocols. In session-level identification, i.e., the protocol identification of the complete traffic from a single website visit, a session consists of multiple flows, so we assign its protocol by majority vote over the flow-level predictions. This aggregation is robust and reduces the session-level error rate to below 0.06\% for every protocol. The high accuracy suggests that the mitigation methods could be applied to the proxied traffic reliably without introducing further distortions the traffic features.

\begin{table}[t]
    \centering
    \caption{Proxy protocol identification results.}
    {\fontsize{8pt}{10pt}\selectfont}
    \begin{tabular}{lcc}
    \toprule
    Protocol 
    & Flow-level Error Rate 
    & Session-level Error Rate \\
    \midrule
    VMess       & 0.255\% (1,127/442,068) & 0.005\% (1/18,207) \\
    Shadowsocks & 0.587\% (2,266/386,017) & 0.000\% (0/16,912) \\
    Trojan      & 0.556\% (2,269/408,340) & 0.059\% (10/16,823) \\
    \bottomrule
    \end{tabular}
    \label{table:proxy_protocol_identification}
\end{table}

\noindent\textbf{Generalization Improvements.} Now we apply \pa for proxy-induced drift to achieve generalization improvement. For comparison, we consider 4 drift mitigating methods adopted in previous works:
\begin{itemize}[leftmargin=*,nosep]
    \item \textit{NetAugment}~\cite{NetCLR23}: It augments the traffic by manipulating the bursts to simulate the traffic from different network environments. It samples packet size and timestamps from historical CDFs to generate more realistic traces;
    \item \textit{DyWin}~\cite{dywin}: DyWin applies window-level stochastic augmentation by masking, jittering, or preserving packet-count windows to simulate burst insertion, merging, and timestamp perturbation in various network conditions;
    \item \textit{Rosetta}~\cite{rosetta}: Rosetta is a TCP-aware augmentation method that simulates transport-level effects such as packet aggregation and packet loss.
    \item \textit{NetRandAugment}~\cite{netrandaugment}: It integrates multiple augmentation operators and samples from them randomly to make the training dataset more diverse for robustness enhancement.
\end{itemize}

We evaluate \pa under different choices of $P_{\text{train}}$ and across multiple WF architectures to examine whether the improvement is tied to a specific training protocol or model design. Table~\ref{table:improvement_for_all_models} shows that \pa consistently improves cross-protocol generalization over the unmitigated baseline for all evaluated WF models and all protocol pairs. The gains are especially pronounced for drift-sensitive models such as DF, BAPM, and NetCLR, while even more drift-tolerant models such as TF and TikTok still benefit from \textsf{PA3}. The standard deviations are generally small, indicating that the improvement is stable across repeated runs.

Compared with existing augmentation-based mitigation methods, \pa achieves the best results in almost all settings, with only one exception for TikTok. This comparison suggests that generic traffic augmentation, which mainly simulates network- or transport-layer variations, is insufficient for proxy-layer drift. In contrast, \pa explicitly uses protocol-specific fingerprints of proxy-induced distortions to align traffic across protocols, which explains its stronger and more consistent generalization performance.

\begin{figure}[t]
    \centering
    \includegraphics[width=.5\textwidth,center]{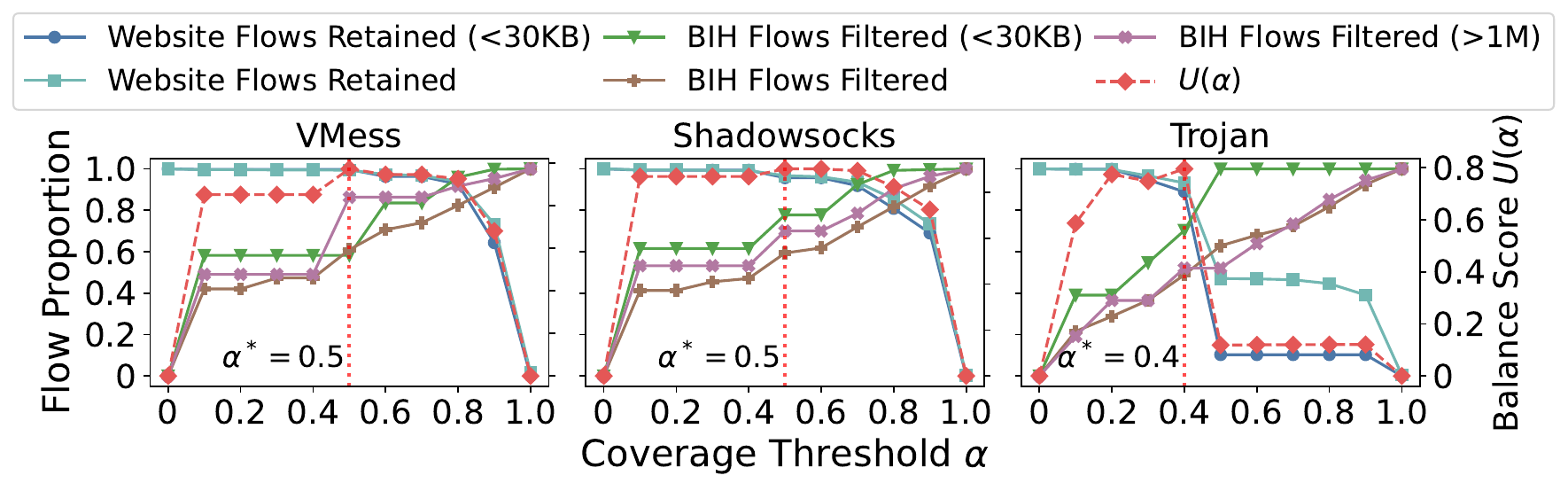}
    \caption{The proportions of BIH flows filtered and website flows retained under different coverage thresholds $\alpha$. We use the balance scoring function $U(\alpha)$ to measure the trade-off between the proportion of filtered BIH flows and that of website flows retained. The optimal $\alpha^*$ is the one that maximizes $U(\alpha)$.}
    \label{fig:coverage}
\end{figure}

\begin{figure}[t]
    \centering
    \includegraphics[width=.5\textwidth, center]{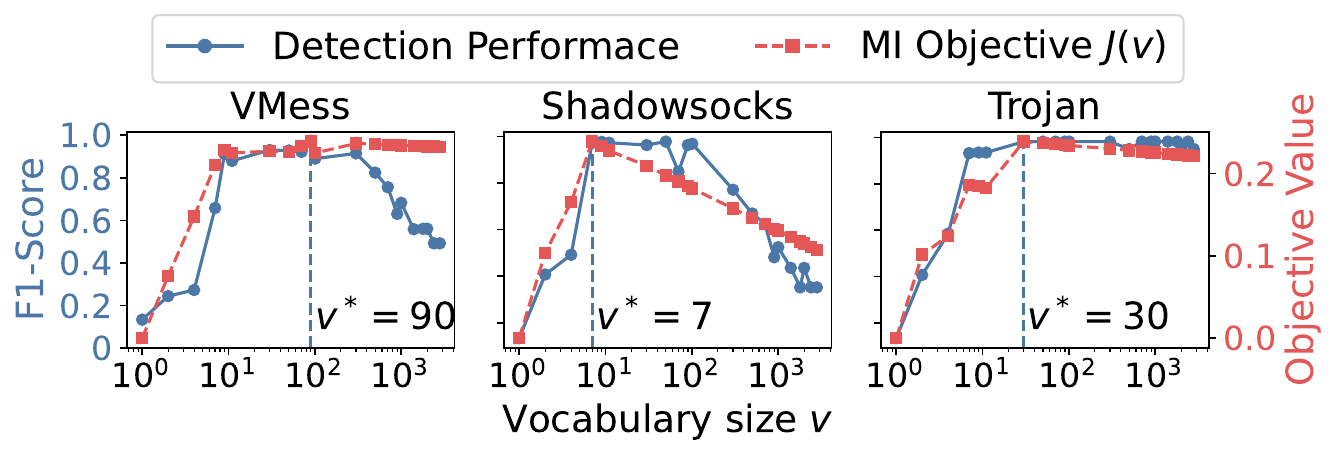}
    \caption{The optimal vocabulary size $|V|^*$ for $n$-gram approach for each proxy protocol. When the MI objective function is optimized, the $n$-gram reaches the best performance.}
    \label{fig:ngram}
\end{figure}

\begin{table*}
    \centering
    \caption{Performance comparison across WF models and mitigation methods, with standard deviations reported for stability. \pa consistently improves cross-protocol generalization and outperforms existing mitigation methods in almost all settings.}
    \begingroup
    \scriptsize
    \setlength{\tabcolsep}{6pt}
    \begin{tabular}{@{\extracolsep{\fill}}c|c|c|cccccc}
    \toprule
    \multirow{2}{*}{$P_{\text{train}}$} & \multirow{2}{*}{$P_{\text{test}}$}  &  \multirow{2}{*}{Mitigation Method} & \multicolumn{6}{c}{WF Attack Model} \\
    \cline{4-9}
     &                                           &                            & DF                        & BAPM                & TF                  & NetCLR              & TikTok              & RF                    \\
    \midrule                          
    \multirow{12}{*}{VMess}             & \multirow{6}{*}{Shadowsocks}        & Baseline                  & 0.518 ($\pm 0.029$) & 0.382 ($\pm 0.007$) & 0.909 ($\pm 0.005$) & 0.332 ($\pm 0.011$) & 0.874 ($\pm 0.007$) & 0.634 ($\pm 0.006$)   \\
                                        &                                     & NetAugment                & 0.563 ($\pm 0.029$) & 0.339 ($\pm 0.008$) & 0.930 ($\pm 0.004$) & 0.430 ($\pm 0.013$) & 0.881 ($\pm 0.007$) & 0.649 ($\pm 0.005$)   \\
                                        &                                     & DyWin                     & 0.587 ($\pm 0.030$) & 0.332 ($\pm 0.012$) & 0.925 ($\pm 0.002$) & 0.449 ($\pm 0.044$) & 0.873 ($\pm 0.007$) & 0.678 ($\pm 0.010$)   \\
                                        &                                     & NetRandAugment            & 0.634 ($\pm 0.015$) & 0.239 ($\pm 0.008$) & 0.922 ($\pm 0.003$) & 0.522 ($\pm 0.027$) & 0.886 ($\pm 0.005$) & 0.644 ($\pm 0.006$)   \\
                                        &                                     & Rosetta                   & 0.589 ($\pm 0.038$) & 0.302 ($\pm 0.011$) & 0.927 ($\pm 0.002$) & 0.508 ($\pm 0.011$) & 0.876 ($\pm 0.006$) & 0.719 ($\pm 0.006$)   \\
                                        &                                     & \pa                       & \underline{\textbf{0.774}} ($\pm 0.008$) & \underline{\textbf{0.653}} ($\pm 0.008$) & \underline{\textbf{0.966}} ($\pm 0.001$) & \underline{\textbf{0.743}} ($\pm 0.008$) & \underline{\textbf{0.927}} ($\pm 0.003$) & \underline{\textbf{0.758}} ($\pm 0.007$)  \\   
    \cline{2-9}
                                        & \multirow{6}{*}{Trojan}             & Baseline                  & 0.481 ($\pm 0.030$) & 0.428 ($\pm 0.007$) & 0.873 ($\pm 0.004$) & 0.389 ($\pm 0.012$) & 0.782 ($\pm 0.005$) & 0.603 ($\pm 0.014$)  \\
                                        &                                     & NetAugment                & 0.514 ($\pm 0.010$) & 0.404 ($\pm 0.008$) & 0.892 ($\pm 0.002$) & 0.429 ($\pm 0.007$) & 0.790 ($\pm 0.010$) & 0.598 ($\pm 0.009$)  \\
                                        &                                     & DyWin                     & 0.544 ($\pm 0.030$) & 0.410 ($\pm 0.009$) & 0.885 ($\pm 0.001$) & 0.455 ($\pm 0.020$) & 0.766 ($\pm 0.007$) & 0.588 ($\pm 0.002$)  \\
                                        &                                     & NetRandAugment            & 0.599 ($\pm 0.017$) & 0.408 ($\pm 0.008$) & 0.874 ($\pm 0.004$) & 0.491 ($\pm 0.007$) & 0.787 ($\pm 0.002$) & 0.552 ($\pm 0.012$)  \\
                                        &                                     & Rosetta                   & 0.428 ($\pm 0.049$) & 0.363 ($\pm 0.014$) & 0.866 ($\pm 0.002$) & 0.328 ($\pm 0.022$) & 0.765 ($\pm 0.007$) & 0.658 ($\pm 0.009$)  \\
                                        &                                     & \pa                       & \underline{\textbf{0.633}} ($\pm 0.011$) & \underline{\textbf{0.582}} ($\pm 0.005$) & \underline{\textbf{0.931}} ($\pm 0.004$) & \underline{\textbf{0.572}} ($\pm 0.009$) & \underline{\textbf{0.830}} ($\pm 0.007$) & \underline{\textbf{0.686}} ($\pm 0.012$)  \\
    \cline{1-9}
    \multirow{12}{*}{Shadowsocks}       & \multirow{6}{*}{VMess}              & Baseline                  & 0.489 ($\pm 0.004$) & 0.360 ($\pm 0.006$) & 0.900 ($\pm 0.002$) & 0.367 ($\pm 0.003$) & 0.688 ($\pm 0.008$) & 0.449 ($\pm 0.012$)   \\
                                        &                                     & NetAugment                & 0.529 ($\pm 0.012$) & 0.428 ($\pm 0.005$) & 0.907 ($\pm 0.002$) & 0.445 ($\pm 0.012$) & 0.690 ($\pm 0.007$) & 0.474 ($\pm 0.011$)   \\
                                        &                                     & DyWin                     & 0.491 ($\pm 0.020$) & 0.438 ($\pm 0.006$) & 0.898 ($\pm 0.002$) & 0.415 ($\pm 0.015$) & 0.632 ($\pm 0.007$) & 0.489 ($\pm 0.010$)   \\
                                        &                                     & NetRandAugment            & 0.555 ($\pm 0.005$) & 0.454 ($\pm 0.012$) & 0.891 ($\pm 0.004$) & 0.492 ($\pm 0.003$) & \underline{\textbf{0.750}} ($\pm 0.007$) & 0.438 ($\pm 0.009$)   \\
                                        &                                     & Rosetta                   & 0.489 ($\pm 0.016$) & 0.439 ($\pm 0.014$) & 0.908 ($\pm 0.004$) & 0.446 ($\pm 0.012$) & 0.662 ($\pm 0.006$) & 0.511 ($\pm 0.012$)   \\
                                        &                                     & \pa                       & \underline{\textbf{0.605}} ($\pm 0.006$) & \underline{\textbf{0.497}} ($\pm 0.006$) & \underline{\textbf{0.930}} ($\pm 0.003$) & \underline{\textbf{0.524}} ($\pm 0.009$) & 0.727 ($\pm 0.006$) & \underline{\textbf{0.580}} ($\pm 0.018$)  \\                              
    \cline{2-9}
                                        & \multirow{6}{*}{Trojan}             & Baseline                  & 0.459 ($\pm 0.014$) & 0.454 ($\pm 0.002$) & 0.876 ($\pm 0.005$) & 0.424 ($\pm 0.001$) & 0.703 ($\pm 0.007$) & 0.491 ($\pm 0.010$)  \\
                                        &                                     & NetAugment                & 0.496 ($\pm 0.016$) & 0.482 ($\pm 0.003$) & 0.882 ($\pm 0.003$) & 0.456 ($\pm 0.005$) & 0.699 ($\pm 0.004$) & 0.502 ($\pm 0.010$)  \\
                                        &                                     & DyWin                     & 0.450 ($\pm 0.030$) & 0.446 ($\pm 0.003$) & 0.866 ($\pm 0.003$) & 0.420 ($\pm 0.016$) & 0.678 ($\pm 0.004$) & 0.501 ($\pm 0.005$)  \\
                                        &                                     & NetRandAugment            & 0.482 ($\pm 0.019$) & 0.510 ($\pm 0.006$) & 0.864 ($\pm 0.001$) & 0.445 ($\pm 0.002$) & 0.744 ($\pm 0.008$) & 0.467 ($\pm 0.019$)  \\
                                        &                                     & Rosetta                   & 0.457 ($\pm 0.007$) & 0.440 ($\pm 0.006$) & 0.873 ($\pm 0.001$) & 0.400 ($\pm 0.005$) & 0.691 ($\pm 0.008$) & 0.534 ($\pm 0.009$)  \\
                                        &                                     & \pa                       & \underline{\textbf{0.600}} ($\pm 0.010$) & \underline{\textbf{0.528}} ($\pm 0.005$) & \underline{\textbf{0.905}} ($\pm 0.001$) & \underline{\textbf{0.518}} ($\pm 0.009$) & \underline{\textbf{0.783}} ($\pm 0.006$) & \underline{\textbf{0.619}} ($\pm 0.004$) \\
    \cline{1-9}
    \multirow{12}{*}{Trojan}            & \multirow{6}{*}{VMess}              & Baseline                  & 0.494 ($\pm 0.008$) & 0.391 ($\pm 0.007$) & 0.902 ($\pm 0.002$) & 0.331 ($\pm 0.008$) & 0.737 ($\pm 0.005$) & 0.593 ($\pm 0.008$)   \\
                                        &                                     & NetAugment                & 0.541 ($\pm 0.022$) & 0.472 ($\pm 0.011$) & 0.919 ($\pm 0.003$) & 0.449 ($\pm 0.012$) & 0.749 ($\pm 0.006$) & 0.579 ($\pm 0.011$)   \\
                                        &                                     & DyWin                     & 0.530 ($\pm 0.029$) & 0.445 ($\pm 0.017$) & 0.894 ($\pm 0.003$) & 0.458 ($\pm 0.017$) & 0.710 ($\pm 0.007$) & 0.575 ($\pm 0.010$)   \\
                                        &                                     & NetRandAugment            & 0.538 ($\pm 0.005$) & 0.527 ($\pm 0.012$) & 0.902 ($\pm 0.002$) & 0.457 ($\pm 0.015$) & 0.747 ($\pm 0.004$) & 0.530 ($\pm 0.010$)   \\
                                        &                                     & Rosetta                   & 0.472 ($\pm 0.016$) & 0.465 ($\pm 0.014$) & 0.891 ($\pm 0.003$) & 0.415 ($\pm 0.017$) & 0.700 ($\pm 0.003$) & 0.590 ($\pm 0.004$)   \\
                                        &                                     & \pa                       & \underline{\textbf{0.606}} ($\pm 0.012$) & \underline{\textbf{0.530}} ($\pm 0.013$) & \underline{\textbf{0.928}} ($\pm 0.003$) & \underline{\textbf{0.538}} ($\pm 0.016$) & \underline{\textbf{0.750}} ($\pm 0.007$) & \underline{\textbf{0.636}} ($\pm 0.011$)\\                           
    \cline{2-9}
                                        & \multirow{6}{*}{Shadowsocks}        & Baseline                  & 0.558 ($\pm 0.043$) & 0.359 ($\pm 0.009$) & 0.922 ($\pm 0.002$) & 0.319 ($\pm 0.013$) & 0.838 ($\pm 0.007$) & 0.667 ($\pm 0.016$)  \\
                                        &                                     & NetAugment                & 0.618 ($\pm 0.032$) & 0.374 ($\pm 0.009$) & 0.941 ($\pm 0.003$) & 0.372 ($\pm 0.024$) & 0.847 ($\pm 0.002$) & 0.646 ($\pm 0.028$)  \\
                                        &                                     & DyWin                     & 0.515 ($\pm 0.058$) & 0.273 ($\pm 0.006$) & 0.923 ($\pm 0.003$) & 0.357 ($\pm 0.031$) & 0.837 ($\pm 0.007$) & 0.654 ($\pm 0.009$)  \\
                                        &                                     & NetRandAugment            & 0.601 ($\pm 0.032$) & 0.335 ($\pm 0.010$) & 0.925 ($\pm 0.004$) & 0.429 ($\pm 0.021$) & 0.844 ($\pm 0.006$) & 0.612 ($\pm 0.017$)  \\
                                        &                                     & Rosetta                   & 0.613 ($\pm 0.009$) & 0.374 ($\pm 0.013$) & 0.927 ($\pm 0.003$) & 0.539 ($\pm 0.025$) & 0.831 ($\pm 0.003$) & 0.651 ($\pm 0.007$)  \\
                                        &                                     & \pa                       & \underline{\textbf{0.691}} ($\pm 0.021$) & \underline{\textbf{0.523}} ($\pm 0.019$) & \underline{\textbf{0.942}} ($\pm 0.003$) & \underline{\textbf{0.576}} ($\pm 0.030$) & \underline{\textbf{0.860}} ($\pm 0.006$) & \underline{\textbf{0.737}} ($\pm 0.006$) \\
    \bottomrule
    \end{tabular}
    \endgroup
    \label{table:improvement_for_all_models}
\end{table*}

\begin{table*}[ht]
    \centering
    {\fontsize{8pt}{10pt}\selectfont}
    \caption{Ablation study of \textsf{PA3}. It demonstrates that each tailoring module mitigates the drift in the corresponding stage.}
    \begin{tabular}{@{\extracolsep{\fill}}cccccccccc}
        \toprule
$P_{\text{train}}$     &           & \multicolumn{2}{c}{VMess} & & \multicolumn{2}{c}{Shadowsocks} & & \multicolumn{2}{c}{Trojan}   \\
 \cline{1-1}\cline{3-4}\cline{6-7}\cline{9-10}
$P_{\text{test}}$    &             & Shadowsocks           & Trojan              &   & VMess                 & Trojan              &          & VMess               & Shadowsocks \\
\midrule
Baseline       &             & 0.518 ($\pm 0.029$)   & 0.481 ($\pm 0.030$) &   & 0.489 ($\pm 0.004$)      & 0.459 ($\pm 0.014$) &          & 0.494 ($\pm 0.008$) & 0.558 ($\pm 0.043$)\\
w/o BFT        &             & 0.726 ($\pm 0.008$)   & 0.600 ($\pm 0.018$) &   & 0.522 ($\pm 0.012$)	  & 0.510 ($\pm 0.011$) &          & 0.601 ($\pm 0.029$) & 0.565 ($\pm 0.039$)\\
w/o PHT        &             & 0.759 ($\pm 0.007$)   & 0.580 ($\pm 0.008$) &   & 0.572 ($\pm 0.021$)	  & 0.514 ($\pm 0.028$) &          & 0.599 ($\pm 0.017$) & 0.670 ($\pm 0.019$)\\
w/o POT        &             & 0.688 ($\pm 0.013$)   & 0.597 ($\pm 0.012$) &   & 0.571 ($\pm 0.010$)	  & 0.574 ($\pm 0.010$) &          & 0.527 ($\pm 0.018$) & 0.599 ($\pm 0.010$)\\
\pa  &             & \underline{\textbf{0.774}} ($\pm 0.008$) & \underline{\textbf{0.633}} ($\pm 0.011$) &   & \underline{\textbf{0.605}} ($\pm 0.006$) & \underline{\textbf{0.600}} ($\pm 0.010$) & & \underline{\textbf{0.606}} ($\pm 0.012$) & \underline{\textbf{0.691}} ($\pm 0.021$)\\
        \bottomrule
    \end{tabular}
    \label{table:ablation}
\end{table*}

\subsection{Parameter Selection from Limited Probe Traffic}
In practice, the amount of probe traffic should be as small as possible. We estimate parameters for the three tailoring modules from a small amount of probe traffic covering only five websites and accounting for 5\% of the complete dataset. Once selected, these parameters are fixed for drift mitigation.

\noindent\textbf{Background Flow Tailoring (BFT).} Fig.~\ref{fig:coverage} reports the empirical threshold search for background flow tailoring. In our experiments, we set $\beta=5$ to prioritize website-flow preservation when computing $U(\alpha)$. The optimal thresholds are $\alpha^*=0.5$ for VMess, $\alpha^*=0.5$ for Shadowsocks, and $\alpha^*=0.4$ for Trojan. Following the alignment rule that we choose the minimal $\alpha$ among all protocols, we select the shared threshold $\alpha_{opt}=0.4$ in this work.

\noindent\textbf{Proxy Handshake Tailoring (PHT).}
The key parameter is the $n$-gram vocabulary size $v$, which controls the granularity of discretization. As defined in Eq.~\ref{eq:mi_objective}, we choose $v$ by maximizing the mutual-information objective $J(v)$, which rewards discriminative anomalous windows and penalizes excessive vocabulary growth. Fig.~\ref{fig:ngram} validates our strategy. The vocabulary size that maximizes $J(v)$ also gives the best F1-score for detecting proxy handshake packets. We therefore use the corresponding $v^*$ to build the anomaly dictionary for each protocol.

\noindent\textbf{Proxy Overhead Tailoring (POT).}
For proxy overhead tailoring, we estimate the Gaussian parameters $\mu$ and $\sigma$ using randomly selected probe websites for each protocol pair. The learned distribution is used to sample the burst-scaling factor $\lambda$. In this evaluation, we sample slope ratios using 5 websites. The parameters for VMess as $P_\text{train}$ are $\mu_\text{VMess}^\text{Shadowsocks}=1.17, \sigma_\text{VMess}^\text{Shadowsocks}=0.15, \mu_\text{VMess}^\text{Trojan}=0.98, \sigma_\text{VMess}^\text{Trojan}=0.12$.

\subsection{Ablation Study}
Now we conduct an ablation study on the 3 traffic tailoring modules within \pa to assess their effects.
\begin{itemize}[leftmargin=*,nosep]
    \item \textit{Background Flow Tailoring (BFT)}: Filter out flows whose sizes lie in the BIH flow size range;
    \item \textit{Proxy Handshake Tailoring (PHT)}: Search for proxy handshake packets and their corresponding ACK packets, and then remove them from each flow;
    \item \textit{Proxy Overhead Tailoring (POT)}: Augment the data transmission part of each flow by rescaling the burst sizes according to the empirical slope ratio range.
\end{itemize}

Table~\ref{table:ablation} shows that every ablated variant improves over the unmitigated baseline in all settings, confirming that each individual module can reduce part of the proxy-induced distortion. Moreover, full \pa consistently achieves the best F1-score across all protocol pairs. The performance improvements are also consistent across protocol pairs. This indicates that the three modules are complementary. BFT removes protocol-specific noise introduced during flow initialization, PHT aligns the extra proxy handshake sequence, and POT compensates for segmentation shifts during content transmission.

\subsection{Mitigation Stability under Limited Probe Traffic}
\label{subsec:limited_data_stability}

In this section, we evaluate the stability of parameter selection when the amount of probe traffic is reduced. The results show that parameters estimated from limited data closely approximate those obtained with sufficient data.

\noindent\textbf{Stability of BFT.} We evaluate how the selected coverage threshold $\alpha^*$ and its corresponding balance score $U(\alpha^*)$ change when fewer BIH flows are available. Since BIH flows are uniformly distributed across websites, we randomly sample BIH flows and reduce the sampling ratio from 20\% to 5\%. The results in Table~\ref{tab:bih_param_stability} show that $U(\alpha^*)$ is stable for all proxy protocols, and $\alpha^*$ remains unchanged as the amount of data is reduced, which suggests that the BIH filtration method is stable even with few probe samples.

\begin{figure}[t]
    \centering
    \includegraphics[width=.51\textwidth, center]{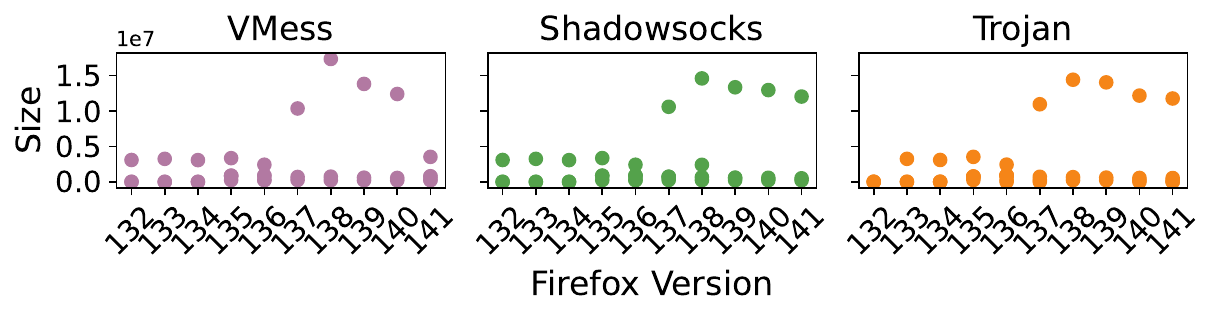}
    \caption{BFT ranges for Firefox versions from 132 to 141.}
    \label{fig:version}
\end{figure}

Fig.~\ref{fig:version} further evaluates stability of the BFT filter ranges under different browser versions. We collect the traffic generated by Firefox version 132 to 141 separately, and compute the ranges for each version across all proxy protocols with $\alpha_{opt}=0.4$. The filter ranges remain stable through version 136 but increase from version 137, mainly because of large \texttt{firefox-settings-attachments.cdn\allowbreak .mozilla.net} flows, which are related to some remote-setting content. Firefox may upgrade its browser configuration policies from version 137. We therefore recommend updating the ranges every 2 or 3 Firefox versions.

\begin{table}[t]
    \centering
    \caption{Stability of the selected optimal coverage $\alpha^*$ and the corresponding balance score $U(\alpha^*)$ range under different sampling ratios.}
    \label{tab:bih_param_stability}
    \small
    \setlength{\tabcolsep}{4pt}
    \begin{tabular}{lcccccc}
    \toprule
    \multirow{2}{*}{Protocol} 
    & \multicolumn{4}{c}{$\alpha^*$ under sampling ratio} 
    & \multirow{2}{*}{$\max \Delta \alpha^*$} 
    & \multirow{2}{*}{$U(\alpha^*)$ range} \\
    \cmidrule(lr){2-5}
    & 20\% & 15\% & 10\% & 5\% &  &  \\
    \midrule
    VMess       & 0.50 & 0.50 & 0.50 & 0.50 & 0.00 & 0.93--0.94 \\
    Shadowsocks & 0.50 & 0.50 & 0.50 & 0.50 & 0.00 & 0.90--0.92 \\
    Trojan      & 0.40 & 0.40 & 0.40 & 0.40 & 0.00 & 0.80--0.83 \\
    \bottomrule
    \end{tabular}
\end{table}

\noindent\textbf{Stability of PHT.} The core parameter in identifying the proxy handshake packet sequence is the $n$-gram vocabulary size $v$, which balances precision and recall. We assess whether $v^*$ can be reliably estimated with limited data by reducing the ratio between the numbers of training and testing samples. The ratio is reduced from $25\%$ down to $5\%$. For each ratio and each protocol, we search for $v^*$ via $J(v)$ to report the best F1-score. Table~\ref{tab:ngram_vocab_stability} shows that the optimal vocabulary size $v^*$ remains unchanged for each protocol, and F1-score stays consistently high ($\ge 0.973$) even when the ratio is only $5\%$. The results indicate the robustness of MI-based vocabulary size selection, and that the proxy handshake patterns are sufficiently regular and distinctive such that limited data suffice to learn an appropriate discretization granularity.

\begin{table}[t]
    \centering
    \caption{Optimal vocabulary size and best F1-score under different training ratios.}
    \label{tab:ngram_vocab_stability}
    \small
    \setlength{\tabcolsep}{4pt}
    \begin{tabular}{lccccc c}
    \toprule
    \multirow{2}{*}{Protocol}
    & \multicolumn{5}{c}{Best F1-score under training ratio}
    & \multirow{2}{*}{$v^*$} \\
    \cmidrule(lr){2-6}
                & 25\%  & 20\%  & 15\%  & 10\%  & 5\%   & \\
    \midrule
    VMess       & 0.973 & 0.976 & 0.975 & 0.975 & 0.976 & 90 \\
    Shadowsocks & 0.977 & 0.979 & 0.978 & 0.980 & 0.977 & 7  \\
    Trojan      & 0.981 & 0.983 & 0.982 & 0.980 & 0.983 & 30 \\
    \bottomrule
    \end{tabular}
\end{table}

\begin{figure}[t]
    \centering
    \includegraphics[width=.5\textwidth, center]{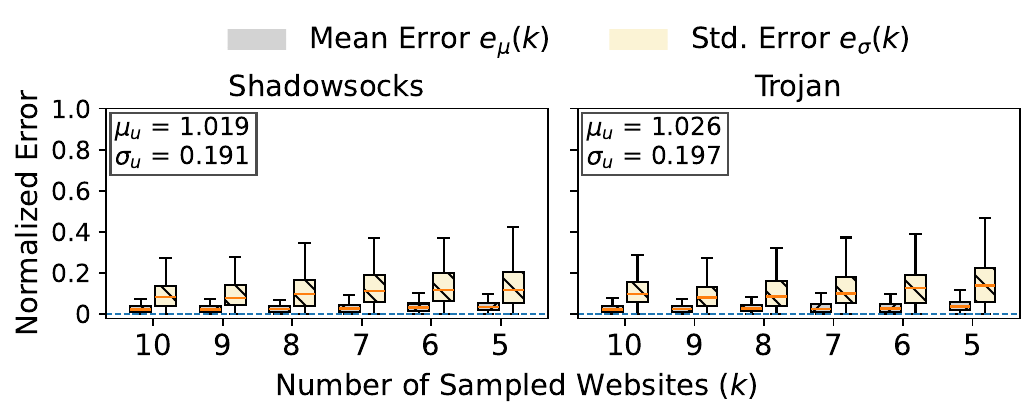}
    \caption{Stability of Gaussian parameter estimation under reduced sampling sizes. VMess is used as $P_\text{train}$. Estimation errors $e_\mu(k), e_\sigma(k)$ are used to measure the effects of target mean ($\mu_u$) and std. ($\sigma_u$) approximation, respectively.}
    \label{fig:gaussian_param_estimation}
\end{figure}

\noindent\textbf{Stability of POT.} We reduce the number of websites within the probe traffic to estimate Gaussian parameters $\mu$ and $\sigma$. Suppose that $\mu_u, \sigma_u$ are target parameters. We use
\begin{equation}
    e_\mu(k)=\frac{|\hat{\mu}_k-\mu_u|}{\mu_u}, \ e_\sigma(k)=\frac{|\hat{\sigma}_k-\sigma_u|}{\sigma_u},
\end{equation}
to measure the normalized estimation error, where $\hat{\mu}_k$ and $\hat{\sigma}_k$ are mean and standard deviation estimates from $k$ websites. Fig.~\ref{fig:gaussian_param_estimation} shows how the errors change when $k$ declines. For each $k$, we repeatedly subsample the candidate websites. Fig.~\ref{fig:gaussian_param_estimation} shows median errors below 0.2 for both protocols, although smaller samples increase the upper-tail variance. Notably, the improvements in Table~\ref{table:improvement_for_all_models} use only $k=5$ probe websites, demonstrating that limited probe traffic remains effective.

\section{Conclusion}
\label{sec:conclusion}
We present the first systematic study of proxy-induced traffic drift in WF, showing that the diversity among proxy protocols leads to poor generalization to unseen protocols. Through payload-aware analysis, we find that proxy protocols introduce distinct distortions throughout a website visit. Based on these findings, we develop \textsf{PA3}, which extracts structural fingerprints from a small amount of decryptable probe traffic and converts them into payload-agnostic rules for tailoring encrypted target traffic. Experiments on real-world proxied traffic show that \pa considerably improves generalization to unseen protocols. These results demonstrate that explainable traffic alignment can substantially improve WF robustness under realistic proxy protocol changes.

\bibliographystyle{IEEEtran}
\bibliography{ref}

\vfill

\end{document}